\documentclass[12pt]{article}
\usepackage{amssymb, amsmath, amsfonts}

\newtheorem{theorem}{Theorem}
\newtheorem{lemma}[theorem]{Lemma}
\newtheorem{definition}[theorem]{Definition}

\newtheorem{example}[theorem]{Example}
\newtheorem{remark}[theorem]{Remark}

\begin{document}

\title{Semi-discrete hyperbolic equations admitting five dimensional characteristic
$x$-ring}
\author{ Kostyantyn Zheltukhin$^1$ and Natalya Zheltukhina$^2$\\
\small $^1$ Department of Mathematics, Faculty of Science,\\
\small Middle East Technical University 06531 Ankara, Turkey.\\
\small email:zheltukh@metu.edu.tr\\
\small $^2$ Department of Mathematics, Faculty of Science,\\
\small  Bilkent University, 06800 Ankara, Turkey.\\
\small email:natalya@fen.bilkent.edu.tr}

\begin{titlepage}

\maketitle

\begin{abstract}
The necessary and sufficient conditions for a hyperbolic semi-discrete equation to
have five dimensional characteristic {\it x}-ring are derived. For any given chain, the derived  conditions are easily verifiable
by straightforward calculations.
\end{abstract}


\medskip


\end{titlepage}

\section{Introduction}

In the present paper we are considering integrability of hyperbolic type semi-discrete equations.
There exist many different approaches to define and classify integrable equations:
symmetry approach, Peinlev\'{e} analysis, method of algebraic entropy and other methods.
For classification of hyperbolic type equations  the approach based on
the notions of characteristic rings turns out to be very effective.

The notion of a characteristic ring was introduced by Shabat  to classify  hyperbolic systems of exponential type
\begin{equation}
u^i_{xy}=e^{(a_{i1}u_1+a_{i2}u_2+ \dots + a_{in}u_n)}\qquad i=1,2, \dots n,
\end{equation}
such system has a finite dimensional characteristic ring if and only if $A=(a_{i,j})$  is a Cartan matrix of a semi-simple Lie algebra, see  \cite{ShYa}.
Then in \cite{LSSh} it was shown that a system of hyperbolic equations
\begin{equation}
u^i_{xy}=f^i(u_1,u_2,\dots u_n)\qquad i=1,2, \dots n
\end{equation}
can be integrated in quadratures if its characteristic ring is finite dimensional.

Zhiber and his collaborators considered application of the characteristic ring to classification problems of general hyperbolic equations
\begin{equation}\label{cont_eqn}
u_{xy}=f(u,u_x,u_y).
\end{equation}
In particular the classification of equations Eq(\ref{cont_eqn}) admitting  two dimensional, three dimensional  or four dimensional (for some special form of function $f$) characteristic rings
was considered in \cite{ZhMur}-\cite{Mur}. For other classification results based on the notion of the characteristic ring see \cite{ZhM}- \cite{SokZhSt} and a review paper \cite{ZhMHSh}.

Later  Habibullin extended  the notion of characteristic ring  to semi-discrete and discrete equations and applied this notion to solve  different classification problems for such equations
(see \cite{H}-\cite{HZh}).

Let us give necessary definitions.
Consider a hyperbolic type semi-discrete equation
\begin {equation} \label{basic_eqn}
t_{1x}=f(x,t,t_1,t_x),
\end{equation}
where the function $t(n,x)$ depends on discrete variable $n$ and continuous variable
$x$. We use the  following notations
$t_x=\frac{\partial }{\partial x} t$, $t_1=t(n+1,x)$, and
$t_{[k]}=\frac{\partial ^k }{\partial x^k} t$, where $k\in {\mathbb N}$
and $t_m=t(n+m,x)$, $m\in {\mathbb Z}$.
\begin{definition}
 A function $F(x,t,t_1,\dots ,t_k)$ is called an $x$-integral of the equation Eq.(\ref{basic_eqn}) if
$$
 D_x F(x,t,t_1,\dots ,t_k)=0
$$
for all solutions of Eq.(\ref{basic_eqn}). The operator $D_x$ is the total derivative with respect to $x$.

\noindent
A function $G(x,t,t_x,\dots ,t_{[m]})$ is called an $n$-integral of the equation Eq.(\ref{basic_eqn}) if
$$
 D G(x,t,t_x,\dots ,t_{[m]})=G(x,t,t_x,\dots ,t_{[m]})
$$
for all solutions of Eq.(\ref{basic_eqn}).
\end{definition}
\noindent
The equation Eq.(\ref{basic_eqn}) is called Darboux integrable if it admits  non trivial
$x$- and $n$- integrals (see \cite{HP}).
\begin{example}
For example the equation
\begin{equation}\label{eq_ex3}
t_{1x}=\frac{tt_1}{t_x}
\end{equation}
 has an $x$-integral $F=\displaystyle{\frac{t_2}{t}}$ and an $n$-integral $I=\displaystyle{\frac{t}{t_x}+\frac{t_x}{t}}$. Hence the equation is Darboux integrable.
\end{example}

We note that a Darboux integrable equation can be reduced to a pair
of ordinary equations: ordinary differential equation and ordinary difference equation.

In \cite{HP}   an effective criterion for the existence of $x$- and $n$-integrals was given.
\begin{theorem} \cite{HP}
 An equation Eq.(\ref{basic_eqn}) admits a non-trivial $x$-integral if and only if its
characteristic x-ring is of finite dimension.\\
 An equation Eq.(\ref{basic_eqn}) admits a non-trivial n-integral if and only if its
characteristic $n$-ring is of finite dimension.
\end{theorem}

It is generally believed that a finite dimensional characteristic $x$-ring can not have
dimension larger than five. The examples of Darboux integrable semi-discrete equations known to us support this hypothesis. On the other hand one can construct examples of Darboux integrable
semi-discrete equations with  characteristic $n$-ring of an arbitrary large finite dimension.  So we study semi-discrete equation Eq.(\ref{basic_eqn}) with five dimensional
characteristic $x$-ring.  The case of three and four dimensional rings  were considered in \cite{HZhP1} and \cite{ZhZh} respectively.

In general it is not easy to determined the dimension of the characteristic ring. In our paper we give the
necessary and sufficient conditions for the characteristic $x$-ring to be five dimensional. The derived conditions are checked by
straightforward calculations and can be effectively used to determine if the  the characteristic $x$-ring is five dimensional.
We also present two examples of equations that have five dimensional characteristic $x$-ring.

The paper is organized as follows. In Section \ref{x-ring_general} we introduce  the characteristic
$x$-ring for a general equation Eq.(\ref{basic_eqn}). In Section \ref{x-ring_5d}  we derive necessary and
sufficient conditions for the characteristic $x$-ring to be five dimensional and give two
example of an equation with five dimensional $x$-ring. Equation Eq.(\ref{eq_ex1}) was introduced in \cite{AS}. The second equation Eq.(\ref{eq_ex2}) we believe to be new.
 Note that equation Eq.(\ref{eq_ex3}) possesses four dimensional characteristic $x$-ring.

\section{Characteristic ring of a hyperbolic type equation}\label{x-ring_general}

The characteristic $x$-ring $L_x$ of the equation Eq.(\ref{basic_eqn}) is generated by two
vector fields (see \cite{HP})
$$
X=\frac{\partial}{\partial t_x}, \qquad
$$
and
$$
K=\frac{\partial }{\partial x}+ t_x \frac{\partial }{\partial t }+f\frac{\partial
}{\partial t_1}+g \frac{\partial }{\partial t_{-1}}+f_1\frac{\partial }{\partial t_2
}+ ...
$$
where function $g$ is determined by
\begin{equation}
t_{-1x}=g(x,t_{-1},t,t_x).
\end{equation}
To obtain above equality we apply $D^{-1}$ to Eq.(\ref{basic_eqn}) and
then solve the resulting equation for $t_{-1x}$.

Let us  introduce some  vector fields from $L_x$.
\begin{equation}
C_1=[X,K]\quad \mbox{and} \quad  C_n=[X,C_{n-1}] \quad n=2,3,\dots
\end{equation}
and
\begin{equation}
Z_1=[K,C_1]\quad \mbox{and} \quad  Z_n=[K,Z_{n-1}] \quad n=2,3,\dots
\end{equation}
To write this vector fields  it is convenient to define the following quantities
\begin{equation}\label{notation}
p=\frac{f_x+t_xf_t+ff_{t_1}}{f_{t_x}}, \quad v=f_t+f_{t_x}f_{t_1}, \quad w=f_{xt_x}
+t_xf_{t_x t}+f f_{t_x t_1}, \quad h=f_{t_xt_xt_x}f_{t_x}-3f_{t_xt_x}^2 .
\end{equation}
We have
$$
\begin{array}{l}
 \displaystyle{C_1=\frac{\partial}{\partial t} + f_{t_x}\frac{\partial}{\partial
t_1}+ g_{t_x}\frac{\partial}{\partial t_{-1}}+\dots}\\
 \\
 \displaystyle{C_2= f_{t_xt_x}\frac{\partial}{\partial t_1}+
g_{t_xt_x}\frac{\partial}{\partial t_{-1}}+\dots}\\
\\
 \displaystyle{Z_1=(w-v)\frac{\partial }{\partial t_1}+.... }\\
\end{array}
$$
and so on.\\
Let us determine what vectors can form a basis of $L_x$ assuming that $dim\, L_x=5$.

First assume that  $f_{t_xt_x}\ne 0$. We have that vector fields $X$, $K$, $C_1$ and
$C_2$ are  linearly independent. Also, as  was shown in \cite{ZhZh}, if $C_3$ and
$Z_1$ belong to the linear span of $X$, $K$, $C_1$ and $C_2$ then $L_x$ is four
dimensional algebra. To have  a five dimensional algebra one of the vectors
$C_3$, $Z_1$ must be linearly independent of $X$, $K$, $C_1$ and $C_2$. Hence if
$f_{t_xt_x}\ne 0$ the five dimensional algebra $L_x$ is generated either by $X$,
$K$, $C_1$, $C_2$ and $Z_1$ or
by $X$, $K$, $C_1$, $C_2$ and $C_3$.

If $f_{t_xt_x}=0$ then $C_n=0$, $n=2,3\dots$ and  the  algebra is spanned by $X$,
$K$, $C_1$, $Z_1$ and $Z_2$.

 To check that a vector admits the expansion with respect to a particular basis we
use the following~Remark.
 \begin{remark}\label{rem1}
One can  check equalities between vector fields using the automorphism
$D(\,\,)D^{-1}$. Direct calculations show that
$$
DXD^{-1}=\frac{1}{f_x}X,\qquad
DKD^{-1}=K-p X.
$$
The images of other vector fields under this automorphism can be obtained by
commuting  $DXD^{-1}$ and $DKD^{-1}$.
\end{remark}

\section{ Five dimensional characteristic $x$-rings}\label{x-ring_5d}

\subsection{Case 1}

 Let us find conditions for the characteristic algebra $L_x$  to be generated by
linearly independent vector fields $X$, $K$, $C_1$, $C_2$ and $Z_1$. (We assume
that $f_{t_xt_x}\ne 0$.)\\

As the next lemma shows to check that  vector fields $X$, $K$, $C_1$, $C_2$ and $Z_1$ form a basis
of  $L_x$ it is enough to check that the vectors fields $C_3$, $[K,C_2]$  and $[K,Z_1]$ have unique
expansions. Also we note that if  $C_3$, $[K,C_2]$  and $[K,Z_1]$ can be expended with respect to
 $X$, $K$, $C_1$, $C_2$ and $Z_1$ then  they are linear combinations of $C_2$ and $Z_1$ only
\begin{equation}\label{case1.1}
C_3 = \alpha C_2 +\beta Z_1,
\end{equation}
\begin{equation}\label{case1.2}
[K,C_2] = \gamma C_2 +\mu Z_1,
\end{equation}
\begin{equation}\label{case1.3}
[K,Z_1] = \eta C_2 +\sigma Z_1,
\end{equation}
for some functions $\alpha,\,\beta,\,\gamma,\,\mu,\,\eta$ and $\sigma$.
This follows from the fact that
$$
X=\frac{\partial}{\partial t_x}, \quad K=\frac{\partial}{\partial x}+ t_x\frac{\partial}{\partial t}+\dots \quad C_1=\frac{\partial}{\partial t}+\dots
$$
but vector fields $C_3$, $[K,C_2]$ and  $[K,Z_1]$ do not contain $\frac{\partial}{\partial t_x}$, $\frac{\partial}{\partial x}$ and $\frac{\partial}{\partial t}$
in their representations.

\begin{lemma}\label{case1_basis_lemma}
 The vector fields $X$, $K$, $C_1$, $C_2$ and $Z_1$ form a basis of the characteristic $x$-ring $L_x$ if and only if vectors fields $C_3$, $[K,C_2]$ and $[K,Z_1]$
admit a unique linear representations with respect to the basis vector fields.
\end{lemma}
\noindent
{\bf Proof.}
 We need to prove that if  vectors fields $C_3$, $[K,C_2]$ and $[K,Z_1]$
admit a unique linear representations with respect to the basis  then all other commutators of the basis vector fields, in particular $[Z_1,X]$, $[C_1,C_2]$, $[C_1,Z_1]$ and $[C_2,Z_1]$,
also admit  unique linear representations. Assume that $C_3$, $[K,C_2]$ and $[K,Z_1]$  have unique linear representation with respect to the basis vector fields.
That is equalities Eq(\ref{case1.1})-Eq(\ref{case1.3}) hold.

Let us show that vector field $[Z_1,X]$,  has unique expansion with respect to the basis.
Using definitions of vector fields $Z_1$, $C_1$, $C_2$ and Jacobi identity  we can write
$$
[Z_1,X]=[[K,C_1],X]=-([[C_1,X],K]+[[X,K],C_1] )=[C_2,K]-[C_1,C_1]=-[K,C_2].
$$
Thus from Eq(\ref{case1.2}) it follows that
\begin{equation}\label{[Z1,X]}
[Z_1,X]=-[K,C_2]=-\gamma C_2 -\mu Z_1.
\end{equation}

Let us show that the vector field $[C_1,C_2]$ has unique expansion with respect to the basis.

Using the definition of vector fields $C_1$, $C_3$ and Jacobi identity  we can write
$$
[C_1,C_2]=[[X,K],C_2]=-([[K,C_2],X]+[[C_2,X],K])=[X,[C_2,K]]+[C_3,K]
$$
Then from Eq(\ref{case1.1}) and Eq(\ref{case1.3}) it follows that
\begin{multline*}
[C_1,C_2]=[X,\gamma C_2+\mu Z_1]+[\alpha C_2+\beta Z_1, K] \\
\shoveleft =X(\gamma)C_2+\gamma C_3+X(\mu)Z_1 +\mu[X,Z_1]\\
-K(\alpha) C_2+\alpha[C_2,K]-K(\beta)Z_1+\beta[Z_1,K]\\
\shoveleft =X(\gamma)C_2+\gamma(\alpha C_2 +\beta Z_1) +X(\mu)Z_1 +\mu(\gamma C_2 +\mu Z_1)\\
-K(\alpha) C_2-\alpha(\gamma C_2 +\mu Z_1)-K(\beta)Z_1-\beta(\eta C_2 +\sigma Z_1).
\end{multline*}
Hence,
\begin{equation}\label{[C1,C2]}
[C_1,C_2]=eC_2+qZ_1,
\end{equation}
where $e=(X(\gamma)+\gamma\alpha +\mu\gamma-K(\alpha)-\alpha\gamma- \beta\eta )$ and \\
 $q=(\gamma\beta +X(\mu)+\mu^2-\alpha\mu- K(\beta)-\beta\sigma )$.

Let us show that the vector field $[C_1,Z_1]$ has unique expansion with respect to the basis.
Using the definition of $C_1$ and Jacobi identity  we can write
$$
[C_1,Z_1]=[[X,K],Z_1]=-([[K,Z_1],X] + [[Z_1,X],K])
$$
Using equalities Eq(\ref{case1.3}) and Eq(\ref{[Z1,X]}) we have
\begin{multline*}
[C_1,Z_1]= [X,\eta C_2+\sigma Z_1]+[\gamma C_2+\mu Z_1, K]\\
\shoveleft =X(\eta)C_2+\eta C_3 + X(\sigma)Z_1+\sigma [X,Z_1]\\
-K(\gamma)C_2+\gamma[C_2,K]-K(\mu)Z_1+\mu[Z_1,K]\\
\shoveleft =X(\eta)C_2+\eta(\alpha C_2 +\beta Z_1)+X(\sigma) Z_1+\sigma(\gamma C_2 +\mu Z_1)\\
-K(\gamma)C_2 -\gamma(\gamma C_2 +\mu Z_1)-K(\mu)Z_1-\mu(\eta C_2 +\sigma Z_1)
\end{multline*}
Hence,
\begin{equation}\label{[C1,Z1]}
[C_1,Z_1]=rC_2+sZ_1,
\end{equation}
where $r=(X(\eta)+\alpha\eta+\gamma\sigma-K(\gamma)-\gamma^2 -\mu\eta) $ and \\
$s=(\alpha\beta +X(\sigma)+\mu\sigma-\gamma\mu- K(\mu)-\mu\sigma)$.

Let us show that the vector field $[C_2,Z_1]$ has unique expansion with respect to the basis.
Using the definition of $C_2$ and Jacobi identity  we can write
$$
[C_2,Z_1]=[[X,C_1],Z_1]=-([[C_1,Z_1],X]+[[Z_1,X],C_1])
$$
Using definition of $C_3$ and equalities Eq(\ref{case1.1}) and Eq(\ref{[Z1,X]})-Eq(\ref{[C1,Z1]}) we have
\begin{multline*}
[C_2,Z_1]= [X,rC_2+sZ_1]+[\gamma C_2+\mu Z_1, C_1]\\
\shoveleft =X(r)C_2+rC_3+X(s)Z_1+s[X,Z_1]\\
-C_1(\gamma)C_2 +\gamma [C_2,C_1]-C_1(\mu)Z_1+\mu[Z_1,C_1]\\
\shoveleft =X(r)C_2+r(\alpha C_2+ \beta Z_1)+X(s)Z_1+s(\gamma C_2+\mu Z_1)\\
-C_1(\gamma)C_2-\gamma(eC_2+qZ_1)-C_1(\mu)Z_1-\mu(rC_2 +sZ_1)
\end{multline*}
Hence,
\begin{equation}\label{[C2,Z1]}
[C_2,Z_1]=mC_2+nZ_1,
\end{equation}
where $m=(X(r)+\alpha r +s\gamma-C_1(\gamma)-\gamma e- \mu r)$ and\\
$n=(r\beta +X(s)+s\mu-\gamma q - C_1(\mu)-\mu s)$.
$\Box$

Now let us find under what conditions the equalities Eq(\ref{case1.1})- Eq(\ref{case1.3}) hold.

\noindent
\begin{remark}
Each of the equalities Eq.(\ref{case1.1}), Eq.(\ref{case1.2}) and Eq.(\ref{case1.3}) leads to
a certain system for the coefficients and one  obtains the coefficients by solving
the corresponding system.
Hence the vector fields $X$, $K$, $C_1$, $C_2$ and $Z_1$ form a basis  if and only
if the solutions of the systems, that  determine coefficients, exist and unique.
\end{remark}

\noindent
This remark holds for other cases as well.

Let us write the systems corresponding to equalities Eq.(\ref{case1.1}),
Eq.(\ref{case1.2}) and Eq.(\ref{case1.3}).
\begin{lemma}
The equality Eq.(\ref{case1.1}) holds if and only if the coefficients $\alpha$ and
$\beta$ satisfy the following system
\begin{equation}\label{case1.1_system}
\begin{array}{lllllllll}
E^{(1)}_{11}\beta &+&E^{(1)}_{12}(D\, \beta)& & & & &=& F^{(1)}_1\\
& & E^{(1)}_{22}(D\,\beta)&+&E^{(1)}_{23}\alpha & +& E^{(1)}_{24}(D\, \alpha)&=&
F^{(1)}_2\\
& &E^{(1)}_{32}(D\,\beta)&+& & & E^{(1)}_{34}(D\, \alpha)&=& F^{(1)}_3\\
\end{array}
\end{equation}
where
$$
E^{(1)}_{11}=\frac{1}{f_{t_x}^2},\quad   E^{(1)}_{12}=-1, \quad F^{1}_1=0,\quad
E^{(1)}_{22}=p,\quad   E^{(1)}_{23}=\frac{1}{f_{t_x}^2},\quad
E^{(1)}_{24}=-\frac{1}{f_{t_x}},\quad
 F^{(1)}_{2}=\frac{3f_{t_xt_x}}{f_{t_x}^3},\quad
 $$
 $$
E^{(1)}_{32}=w-v-pf_{t_xt_x},\quad
 E^{(1)}_{34}=\frac{f_{t_xt_x}}{f_{t_x}},\quad     F^{(1)}_{3}=\frac{h}{f_{t_x}^3}.
$$
\end{lemma}
{\bf Proof.}
Applying the automorphism $D(\cdot)D^{-1}$ to Eq.(\ref{case1.1}) we get
\begin{equation}\label{DC3}
DC_3D^{-1}=(D\alpha)DC_2D^{-1}+(D\beta)DZ_1D^{-1}.
\end{equation}
Direct calculations show that $$
DC_2D^{-1}=\frac{1}{f^2_{t_x}}C_2-\frac{f_{t_xt_x}}{f^3_{t_x}}C_1+\frac{f_{t_xt_x}f_{t}}{f^4_{t_x}}
X,
$$
$$
DC_3D^{-1}=\frac{1}{f^3_{t_x}} C_3
-\frac{3f_{t_xt_x}}{f^4_{t_x}}C_2-\frac{h}{f_{t_x}^5}\left( C_1-\frac{f_t}{f_{t_x}}
X \right),
$$
$$
DZ_1D^{-1}=\frac{1}{f_{t_x}}Z_1-\frac{p}{f_{t_x}}C_2+\left(\frac{v-w+pf_{t_xt_x}}{f^2_{t_x}}\right)\left(
C_1-\frac{f_t}{f_{t_x}} X \right).
$$
Substituting these expressions for $DC_3D^{-1}$, $DC_2D^{-1}$, $DZ_1D^{-1}$ into
Eq.(\ref{DC3}) and comparing coefficients of  $C_1$, $C_2$ and $Z_1$ we obtain
Eq.(\ref{case1.1_system}). $\Box$

\begin{lemma}
The equality Eq.(\ref{case1.2}) holds if and only if the coefficients $\gamma$ and
$\mu$ satisfy the following system
\begin{equation}\label{case1.2_system}
\begin{array}{lllllllll}
E^{(2)}_{11}\mu &+ & E^{(2)}_{12}(D\, \mu) & & & & & =& F^{(2)}_1\\
& & E^{(2)}_{22}(D\,\mu)&+&E^{(2)}_{23}\gamma &+& E^{(2)}_{24}(D\, \gamma)&=&
F^{(2)}_2\\
& & E^{(2)}_{32}(D\,\mu)&+& & & E^{(2)}_{34}(D\, \gamma)&=& F^{(2)}_3\\
\end{array}
\end{equation}
where
$$
E^{(2)}_{11}=\frac{1}{f_{t_x}},\quad   E^{(2)}_{12}=-1,\quad
F^{(2)}_{1}=\frac{f_{t_xt_x}}{f_{t_x}^2}+\frac{p\beta}{f_{t_x}},\quad
E^{(2)}_{22}=p, \quad   E^{(2)}_{23}=\frac{1}{f_{t_x}},\quad $$
$$  E^{(2)}_{24}=-\frac{1}{f_{t_x}},\quad
 F^{(2)}_{2}=\frac{2w-p\left(3f_{t_xt_x}-f_{t_x}\alpha \right)}{f_{t_x}^2}, \quad
E^{(2)}_{32}=w-v-f_{t_xt_x}p\quad   E^{(2)}_{34}=\frac{f_{t_xt_x}}{f_{t_x}},
$$

$$
F^{(2)}_{3}= \frac{-3f_{t_xt_x}w+
 f_tf_{t_xt_x}-ph}{f_{t_x}^2}+\frac{f_{xt_xt_x}+
 t_xf_{tt_xt_x}+ff_{t_xt_xt_1}}{f_{t_x}}
$$
\end{lemma}
{\bf Proof.}
Applying the automorphism $D(\cdot)D^{-1}$ to Eq.(\ref{case1.2}) we get
\begin{equation}\label{DKC2}
D[K,C_2]D^{-1}=(D\gamma)DC_2D^{-1}+(D\mu)DZ_1D^{-1}
\end{equation}
Direct calculations show that
$$
D[K,C_2]D^{-1}=\frac{1}{f_{t_x}^2}[K,C_2]+\frac{3pf_{t_xt_x}-2w}{f_{t_x}^3}C_2-\frac{p}{f_{t_x}^2}C_3-\frac{f_{t_xt_x}}{f_{t_x}^3}Z_1
$$
$$
+\left(
\frac{3f_{t_xt_x}w-f_{t_xt_x}f_t+ph}{f_{t_x}^4}-\frac{f_{xt_xt_x}+t_xf_{tt_xt_x}+ff_{t_1t_xt_x}}{f_{t_x}^3}\right)C_1+...X
$$
Substituting the expressions for $D[K,C_2]D^{-1}$, $DC_2D^{-1}$, $DZ_1D^{-1}$ into
Eq.(\ref{DKC2}) and comparing coefficients of  $C_1$, $C_2$ and $Z_1$ we obtain
Eq.(\ref{case1.2_system}). $\Box$

\begin{lemma}
The equality Eq.(\ref{case1.3}) holds if and only if the coefficients $\eta$ and
$\sigma$ satisfy the following system
\begin{equation}\label{case1.3_system}
\begin{array}{lllllllll}
E^{(3)}_{11}\sigma &+ & E^{(3)}_{12}(D\, \sigma)& & & & &=& F^{(3)}_1\\
& & E^{(3)}_{22}(D\,\sigma)&+&E^{(3)}_{23}\eta & +&  E^{(3)}_{24}(D\, \eta)&=&
F^{(3)}_2\\
& & E^{(3)}_{32}(D\,\sigma)&+& & & E^{(3)}_{34}(D\, \eta)&=& F^{(3)}_3\\
\end{array}
\end{equation}
where
$$
E^{(3)}_{11}=1, \quad   E^{(3)}_{12}=-1,\quad
F^{(3)}_{1}=p\left(2\mu -p\beta -\frac{2f_{t_xt_x}}{f_{t_x}}
\right)+\frac{2w-v}{f_{t_x}},
$$
$$
E^{(3)}_{22}=p, \quad   E^{(3)}_{23}=1,\quad   E^{(3)}_{24}=-\frac{1}{f_{t_x}},
$$
$$
 F^{(3)}_{2}=p\left( 2\gamma-p\alpha \right) +
 K(p)+\frac{3p^2f_{t_xt_x}-3pw}{f_{t_x}},
\quad
E^{(3)}_{32}=-f_{t_xt_x}p-v+w, \quad
E^{(3)}_{34}=\frac{f_{t_xt_x}}{f_{t_x}},
$$
$$
F^{(3)}_{3}=-(K-pX)\{f_{t_xt_x}p+v-w\}+(f_{t_xt_x}p+v-w)\frac{2w-f_t-2pf_{t_xt_x}}{f_{t_x}}
$$
\end{lemma}
{\bf Proof.}
Applying the automorphism $D(\cdot)D^{-1}$ to Eq.(\ref{case1.3}) we get
$$
D[K,Z_1]D^{-1}=(D\eta)DC_2D^{-1}+(D\sigma)DZ_1D^{-1}
$$
Direct calculations show that
$$
D[K, Z_1]D^{-1}=\left(\frac{\sigma-2p\mu+p^2\beta
}{f_{t_x}}+\frac{2pf_{t_xt_x}+v-2w}{f^2_{t_x}}\right)Z_1
$$
$$+
\left(-(K-pX)\left\{\frac{p}{f_{t_x}}\right\}+\frac{\eta-2p\gamma+p^2\alpha }{f_{t_x}}-
\frac{p(pf_{t_xt_x}+v-w)}{f^2_{t_x}}\right)C_2
$$
$$
+\left((K-pX)\left\{\frac{pf_{t_xt_x}+v-w}{f^2_{t_x}}\right\}+\frac{f_t(pf_{t_xt_x}+v-w)}{f^3_{t_x}}\right)C_1+...
X
$$
Substituting the expressions for $D[K,Z_1]D^{-1}$, $DC_2D^{-1}$, $DZ_1D^{-1}$ and
comparing coefficients of  $C_1$, $C_2$ and $Z_1$ we obtain Eq.(\ref{case1.3_system}).
$\Box$

All the systems in the above lemmas have similar form, in particular,

\begin{equation}\label{gen_system}
\begin{array}{lllllllll}
E_{11}u &+&E_{12}(D\, u)& & & & &=& F_1\\
& & E_{22}(D\,u)&+&E_{23}v & +& E_{24}(D\, v)&=& F_2\\
& &E_{32}(D\,u)&+& & & E_{34}(D\, v)&=& F_3\\
\end{array}
\end{equation}
where $u$, $v$ are unknowns.

We need conditions for existence of a  unique  solution for such systems. The
conditions are given in the following lemma.

\begin{lemma}\label{gen_system_lem}
 The system Eq.(\ref{gen_system}) has a unique solution if $E_{11}$, $E_{12}$,
$E_{22}$, $E_{23}$, $E_{24}$, $E_{32}$, $E_{34}$ and $F_1$, $F_2$, $F_3$ satisfy
 \begin{equation}\label{gen_system_cond1}
\left(-E_{11}E_{22}E_{34}(D^{-1}E_{34})+E_{11}E_{24}E_{32}(D^{-1}E_{34})-
E_{12}E_{23}E_{34}(D^{-1}E_{32})\right)\ne 0
 \end{equation}
 and
 \begin{equation}\label{gen_system_cond2}
 (D\,H)=\frac{F_1}{E_{12}}-\frac{E_{11}}{E_{12}}H,
 \end{equation}
 where
 \begin{multline}
 H=((F_1E_{24}E_{32}-F_1E_{22}E_{34}-F_2E_{12}E_{34} -F_3E_{12}E_{24})(D^{-1}E_{34})
+(D^{-1}F_{3})E_{12}E_{23}E_{34}) \\
(-E_{11}E_{22}E_{34}(D^{-1}E_{34})+E_{11}E_{24}E_{32}(D^{-1}E_{34})-
E_{12}E_{23}E_{34}(D^{-1}E_{32}))^{-1}
 \end{multline}
 \end{lemma}
{\bf Proof.}
In the  system Eq.(\ref{gen_system}) the coefficients and variables depend on the
discrete variable $n\in { \mathbb Z}$. So we can rewrite the system as follows

\begin{equation}\label{gen_system_n}
\begin{array}{lllllllll}
E_{11}(n)u(n) &+&E_{12}(n)u(n+1)& & & & &=& F_1(n)\\
& & E_{22}(n)u(n+1)&+&E_{23}(n)v(n) & +& E_{24}(n)v(n+1)&=& F_2(n)\\
& &E_{32}(n)u(n+1)&+& & & E_{34}(n) v(n+1)&=& F_3(n)\\
\end{array}
\end{equation}
The above equalities must hold for all values of $n$.
Applying $D^{-1}$ to the last equation above we obtain
$$
E_{32}(n-1)u(n)+E_{34}(n-1)v(n)= F_3(n-1).
$$
Now we have a linear system  to find $u(n),\, v(n),\, u(n+1)$ and $v(n+1)$
independently. The system has a unique solution if condition
Eq.(\ref{gen_system_cond1}) holds. Solving the system we find
\begin{equation}
u(n)=H,\,\, u(n+1)=\frac{F_1}{E_{12}}-\frac{E_{11}}{E_{12}}H
\end{equation}
and
\begin{equation}
v(n)=\frac{(D^{-1}F_3)}{(D^{-1}E_{34})}-\frac{(D^{-1}E_{32})}{(D^{-1}E_{34})}H,\,\,
v(n+1)=\frac{F_3}{E_{34}}-\frac{E_{32}F_1}{E_{34}E_{12}}
+\frac{E_{32}E_{11}}{E_{34}E_{12}}H
\end{equation}

The condition Eq.(\ref{gen_system_cond1}) shows that $D\,u(n)=u(n+1)$ and
$D\,v(n)=v(n+1)$. Hence the system Eq.(\ref{gen_system_n}) has a unique solution.
$\Box$

Now we can give necessary and sufficient conditions for the algebra to be generated
by vector fields $X$, $K$, $C_1$, $C_2$ and $Z_1$.

\begin{theorem}\label{case1_th}
The characteristic $x$-ring of  Eq.(\ref{basic_eqn}) is generated by vector
fields $X$, $K$, $C_1$, $C_2$ and $Z_1$ if and only if the following conditions are
satisfied

 \begin{equation}\label{basic_system_cond1_case1}
 \left(-E_{11}^{(i)}E_{22}^{(i)}E_{34}^{(i)}(D^{-1}E_{34}^{(i)})+E_{11}^{(i)}E_{24}^{(i)}E_{32}^{(i)}(D^{-1}E_{34}^{(i)})-
E_{12}^{(i)}E_{23}^{(i)}E_{34}^{(i)}(D^{-1}E_{32}^{(i)})\right)\ne 0
 \end{equation}
and
 \begin{equation}\label{basic_system_cond2_case1}
 (D\,H^{(i)})=\frac{F^{(i)}_1}{E^{(i)}_{12}}-\frac{E^{(i)}_{11}}{E^{(i)}_{12}}H^{(i)},
 \end{equation}
 where
 \begin{multline}
 H^{(i)}=\left(\left(F_1^{(i)}E_{24}^{(i)}E_{32}^{(i)}-F_1^{(i)}E_{22}^{(i)}E_{34}^{(i)}-F_2^{(i)}E_{12}^{(i)}E_{34}^{(i)}
-F_3^{(i)}E_{12}^{(i)}E_{24}^{(i)}\right)(D^{-1}E_{34}^{(i)})
+(D^{-1}F_{3}^{(i)})E_{12}^{(i)}E_{23}^{(i)}E_{34}^{(i)}\right) \\
\left(-E_{11}^{(i)}E_{22}^{(i)}E_{34}^{(i)}(D^{-1}E_{34}^{(i)})+E_{11}^{(i)}E_{24}^{(i)}E_{32}^{(i)}(D^{-1}E_{34}^{(i)})-
E_{12}^{(i)}E_{23}^{(i)}E_{34}^{(i)}(D^{-1}E_{32}^{(i)})\right)^{-1}
 \end{multline}

where $i=1,2,3$.
\end{theorem}
{\bf Proof.} By Lemma \ref{gen_system_lem} the conditions
Eq.(\ref{basic_system_cond1_case1}), Eq.(\ref{basic_system_cond2_case1})  imply that the
systems Eq.(\ref{case1.1_system}), Eq.(\ref{case1.2_system}) and Eq.(\ref{case1.3_system})
have unique solutions. Hence equalities Eq.(\ref{case1.1}), Eq.(\ref{case1.2}) and
Eq.(\ref{case1.3}) hold and the characteristic ring $L_x$ is generated by vector fields
$X$, $K$, $C_1$, $C_2$ and $Z_1$. $\Box$

\begin{example}\label{ex1}
Consider an equation
\begin{equation}\label{eq_ex1}
t_{1x}t_x=t+t_1
\end{equation}
introduced by  Adler and Startsev  in \cite{AS}.  For this equation one can easily check that the conditions of the Theorem \ref{case1_th} are satisfied. Hence the characteristic ring $L_x$ is five dimensional and generated by vector fields $X$, $K$, $C_1$, $C_2$ and $Z_1$. We have
\begin{equation}
C_3 = -\frac{3}{t_x} C_2, \quad [K,C_2] = -\frac{1}{t_x} Z_1, \quad [K,Z_1] = -\frac{1}{t_x} Z_1.
\end{equation}
The $x$-integral and $n$-integral for the above equation are
$$
F=\frac{(u_3-u_1)(u_2-u)}{(u_2+u_1)}\, , \qquad I=\frac{(u_{xx}-1)^2}{u_x^2}\, .
$$
\end{example}

\begin{example}\label{ex2}
Consider an equation
\begin{equation}\label{eq_ex2}
t_{1x}=\cosh(t_1-t)t_x+\sinh(t_1-t)\sqrt{t_x^2-1}
\end{equation}
For this equation one can easily check that the conditions of the Theorem \ref{case1_th} are satisfied. Hence the characteristic ring $L_x$ is five dimensional
and generated by vector fields $X$, $K$, $C_1$, $C_2$ and $Z_1$. We have
\begin{equation}
C_3 = -\frac{3t_x}{t_x^2-1} C_2, \quad [K,C_2] = -\frac{t_x}{t_x^2-1} Z_1, \quad [K,Z_1] = (t_x^2-1)^{\frac{1}{2}} Z_1.
\end{equation}
The $x$-integral and $n$-integral for the above equation are
$$
\hat{F}=\frac{(e^{t_2}-e^{t_1})(e^{t_3}-e^{t})}{(e^{t_2}-e^{t})(e^{t_3}-e^{t_1})}\, , \qquad \hat{I}=e^{-t}\left(t_x+\sqrt{t_x^2-1}\right)\,.
$$
\end{example}

\subsection{Case 2}

 Let us find conditions for the characteristic algebra $L_x$  to be generated by
vector fields $X$, $K$, $C_1$, $C_2$ and $C_3$. (We assume that $f_{t_xt_x}\ne 0$.)

As the next lemma shows to check that  vector fields $X$, $K$, $C_1$, $C_2$ and $C_3$ form a basis
of  $L_x$ it is enough to check that the vectors fields $Z_1$, $[C_1,C_2]$  and $C_4$ have unique
expansions. Also we note that if $Z_1$, $[C_1,C_2]$  and $C_4$  can be expended with respect to
 $X$, $K$, $C_1$, $C_2$ and $C_3$ then
\begin{equation}\label{case2.1}
Z_1=\tilde \lambda C_2,
\end{equation}
\begin{equation}\label{case2.2}
[C_1,C_2]=\tilde \alpha C_2+\tilde \beta C_3,
\end{equation}
\begin{equation}\label{case2.3}
C_4=\tilde \mu C_2+\tilde \eta C_3.
\end{equation}
 for some functions $\tilde\lambda,\,\tilde\alpha,\,\tilde\beta,\, \tilde\mu$ and $\tilde\eta$.  This follows from the form of $Z_1$, $[C_1,C_2]$  and $C_4$.
Note that if $Z_1=\tilde\lambda_1 C_2+\tilde\lambda_2 C_3$ with $\tilde\lambda_2\ne 0$ we have the Case 1.

\begin{lemma}
The vector fields $X$, $K$, $C_1$, $C_2$ and $C_3$ form a basis of the characteristic $x$-ring $L_x$ if and only if vectors fields $Z_1$, $[C_1,C_2]$ and $C_4$
admit  unique linear representations with respect to the basis vector fields.
\end{lemma}

\noindent
The above Lemma is proved in the same way as Lemma \ref{case1_basis_lemma}.

Let us write the systems corresponding to equalities  Eq.(\ref{case2.2}) and
(\ref{case2.3}). The condition for the equality Eq.(\ref{case2.1}) was obtained in
\cite{ZhZh}.

\begin{lemma}
The equality Eq.(\ref{case2.2}) holds if and only if the coefficients $\tilde \alpha$
and $\tilde \beta$ satisfy the following system
\begin{equation}\label{case2.2_system}
\begin{array}{lllllllll}
\tilde E^{(2)}_{11}\tilde \beta &+ & \tilde E^{(2)}_{12}(D\, \tilde\beta)& & & & &=&
\tilde F^{(2)}_1\\
& & \tilde E^{(2)}_{22}(D\,\tilde\beta)&+&\tilde E^{(2)}_{23}\tilde\alpha & +&
E^{(2)}_{24}(D\, \tilde\alpha)&=& \tilde F^{(2)}_2\\
& & \tilde E^{(2)}_{32}(D\,\tilde\beta)&+& & & \tilde E^{(2)}_{34}(D\,
\tilde\alpha)&=& \tilde F^{(2)}_3\\
\end{array}
\end{equation}
where
$$
\tilde E^{(2)}_{11}=1, \quad   \tilde E^{(2)}_{12}=-1,   \quad \tilde
F^{(2)}_{1}=\frac{v}{f_{t_x}},
$$
$$
\tilde E^{(2)}_{22}=\frac{3f_{t_xt_x}}{f_{t_x}^2},  \quad   \tilde
E^{(2)}_{23}=\frac{1}{f_{t_x}},\quad  \tilde  E^{(2)}_{24}=-1,\quad
 \tilde F^{(2)}_{2}= \frac{2(f_{tt_x}+f_{t_x}f_{t_1t_x})}{f_{t_x}^2}-
 \frac{3f_{t_xt_x}v-f_{t_xt_x}f_t}{f_{t_x}^3},
$$
$$
\tilde E^{(2)}_{32}=\frac{h}{f_{t_x}^2},  \quad   \tilde E^{(2)}_{34}=f_{t_xt_x},\quad
\tilde
F^{(2)}_{3}=\frac{f_{tt_xt_x}+f_{t_x}f_{t_1t_xt_x}-3f_{t_xt_x}f_{t_1t_x}}{f_{t_x}}-\frac{2f_{t_xt_x}f_{tt_x}}{f_{t_x}^2}-
\frac{f_{t_xt_x}^2f_t+vh}{f_{t_x}^3}.
$$
\end{lemma}
{\bf Proof.}
Applying the automorphism $D(\cdot)D^{-1}$ to Eq.(\ref{case2.2}) we get
$$
D[C_1,C_2]D^{-1}=(D\tilde \alpha)D C_2D^{-1}+(D\tilde \beta)D C_3 D^{-1}.
$$
Direct calculations show that
$$
D[C_1,C_2]D^{-1}=\frac{1}{f_{t_x}^3}[C_1,C_2]-\frac{v}{f_{t_x}^4}C_3+\left(-\frac{2(f_{tt_x}+f_{t_x}f_{t_1t_x})}{f_{t_x}^4}-\frac{f_{t_xt_x}f_t}{f_{t_x}^5}+\frac{3vf_{t_xt_x}}{f_{t_x}^5}\right)C_2
$$
$$
+\left(
\frac{f_{t_xt_x}f_{t_1t_x}}{f_{t_x}^4}-\frac{1}{f_{t_x}}C_1\left(\frac{f_{t_xt_x}}{f_{t_x}^3}\right)-\frac{f_{t_xt_x}(f_{tt_x}+f_{t_x}f_{t_1t_x})}{f_{t_x}^5}
+\frac{f_{t_xt_x}^2f_t}{f_{t_x}^6}+\frac{vh}{f_{t_x}^6}\right)C_1 + ... X
$$
Substituting the expressions for $D[C_1,C_2]D^{-1}$, $DC_2D^{-1}$, $DC_3D^{-1}$ and
comparing coefficients of  $C_1$, $C_2$ and $C_3$ we obtain Eq.(\ref{case2.2_system}).
$\Box$

\begin{lemma}
The equality Eq.(\ref{case2.3}) holds if and only if the coefficients $\tilde \mu$ and
$\tilde \eta$ satisfy the following system
\begin{equation}\label{case2.3_system}
\begin{array}{lllllllll}
\tilde E^{(3)}_{11}\tilde \eta &+ & \tilde E^{(3)}_{12}(D\, \tilde\eta)& & & & &=&
\tilde F^{(3)}_1\\
& & \tilde E^{(3)}_{22}(D\,\tilde\eta)&+&\tilde E^{(3)}_{23}\tilde\mu & +& \tilde
E^{(3)}_{24}(D\, \tilde\mu)&=&\tilde F^{(3)}_2\\
& &\tilde E^{(3)}_{32}(D\,\tilde\eta)&+& & &\tilde E^{(3)}_{34}(D\,
\tilde\mu)&=&\tilde F^{(3)}_3\\
\end{array}
\end{equation}
where
$$
\tilde E^{(3)}_{11}=\frac{1}{f_{t_x}}, \quad   \tilde E^{(3)}_{12}=-1,   \quad
\tilde F^{(3)}_{1}=\frac{6f_{t_xt_x}}{f_{t_x}^2},
$$
$$
\tilde E^{(3)}_{22}=\frac{3f_{t_xt_x}}{f_{t_x}^2},  \quad   \tilde
E^{(3)}_{23}=\frac{1}{f_{t_x}^2},\quad  \tilde  E^{(3)}_{24}=-1,\quad
 \tilde F^{(3)}_{2}=\frac{4h-3f_{t_xt_x}^2}{f_{t_x}^4},
$$
$$
\tilde E^{(3)}_{32}=\frac{h}{f_{t_x}^2},  \quad   \tilde E^{(3)}_{34}=f_{t_xt_x},\quad
\tilde
F^{(2)}_{3}=\frac{f_{t_xt_xt_xt_x}f_{t_x}-5f_{t_xt_x}f_{t_xt_xt_x}}{f_{t_x}^3}
-\frac{5f_{t_xt_x}h}{f_{t_x}^4}.
$$
\end{lemma}
{\bf Proof.}
Applying the automorphism $D(\cdot)D^{-1}$ to Eq.(\ref{case2.3}) we get
$$
DC_4D^{-1}=(D\tilde \mu)D C_2D^{-1}+(D\tilde \eta)D C_3 D^{-1}.
$$
Direct calculations show that
$$
DC_4D^{-1}=\frac{1}{f_{t_x}^4}C_4-\frac{6f_{t_xt_x}}{f_{t_x}^5}C_3-\left(\frac{3(f_{t_xt_xt_x}f_{t_x}-4f_{t_xt_x}^2)}{f_{t_x}^6}+\frac{h}{f_{t_x}^6}\right)C_2
-\frac{1}{f_{t_x}}X\left(\frac{h}{f_{t_x}^5}\right)C_1+ ... X.$$
Substituting the expressions for $DC_4D^{-1}$, $DC_2D^{-1}$, $DC_3D^{-1}$ and
comparing coefficients of $C_1$, $C_2$ and $C_3$ we obtain Eq.(\ref{case2.3_system}).
$\Box$

\begin{theorem}
The characteristic $x$-ring of Eq.\ref{basic_eqn} is generated by vector
fields $X$, $K$, $C_1$, $C_2$ and $C_3$ if and only if the following conditions are
satisfied
\begin{equation}\label{basic_system_cond1_case2}
D
\left(\frac{f_{t_xt_xt_x}}{f_{t_xt_x}}\right)=\frac{f_{t_xt_xt_x}f_{t_x}-3f^2_{t_xt_x}}{f_{t_xt_x}f^2_{t_x}},
\end{equation}
\begin{equation}\label{basic_system_cond2_case2}
 \left(-\tilde E_{11}^{(i)}\tilde E_{22}^{(i)}\tilde E_{34}^{(i)}(D^{-1}\tilde
E_{34}^{(i)})+\tilde E_{11}^{(i)}\tilde E_{24}^{(i)}\tilde
E_{32}^{(i)}(D^{-1}\tilde E_{34}^{(i)})- E_{12}^{(i)}\tilde \tilde
E_{23}^{(i)}\tilde E_{34}^{(i)}(D^{-1}\tilde E_{32}^{(i)})\right)\ne 0
 \end{equation}
and
 \begin{equation}\label{basic_system_cond3_case2}
 (D\,\tilde H^{(i)})=\frac{\tilde F^{(i)}_1}{\tilde E^{(i)}_{12}}-\frac{\tilde
E^{(i)}_{11}}{\tilde E^{(i)}_{12}}\tilde H^{(i)},
 \end{equation}
 where
 \begin{multline}
 \tilde H^{(i)}=\left(\left(\tilde F_1^{(i)}\tilde E_{24}^{(i)}\tilde
E_{32}^{(i)}-\tilde F_1^{(i)}\tilde E_{22}^{(i)}\tilde E_{34}^{(i)}-\tilde
F_2^{(i)}\tilde E_{12}^{(i)}\tilde E_{34}^{(i)} -\tilde F_3^{(i)}\tilde
E_{12}^{(i)}\tilde E_{24}^{(i)}\right)(D^{-1}\tilde E_{34}^{(i)}) +(D^{-1}\tilde
F_{3}^{(i)})\tilde E_{12}^{(i)}\tilde E_{23}^{(i)}\tilde E_{34}^{(i)}\right) \\
\left(-\tilde E_{11}^{(i)}\tilde E_{22}^{(i)}\tilde E_{34}^{(i)}(D^{-1}\tilde
E_{34}^{(i)})+\tilde E_{11}^{(i)}\tilde E_{24}^{(i)}\tilde E_{32}^{(i)}(D^{-1}\tilde
E_{34}^{(i)})- E_{12}^{(i)}\tilde \tilde E_{23}^{(i)}\tilde
E_{34}^{(i)}(D^{-1}\tilde E_{32}^{(i)})\right)^{-1}
 \end{multline}
for $i=2,3$.
\end{theorem}
{\bf Proof.} The condition Eq.\ref{basic_system_cond1_case2} implies that the equality
Eq.(\ref{case2.1}) holds , see \cite{ZhZh}. By Lemma \ref{gen_system_lem} the
conditions Eq.(\ref{basic_system_cond2_case2}) and Eq.(\ref{basic_system_cond3_case2})
imply that the systems Eq.(\ref{case2.2_system}) and Eq.(\ref{case2.3_system}) have unique
solutions.
Hence equalities Eq.(\ref{case2.1}), Eq.(\ref{case2.2}) and Eq.(\ref{case2.3}) hold and the
characteristic ring $L_x$ is generated by vector fields $X$, $K$, $C_1$, $C_1$ and
$C_3$. $\Box$

\subsection{Case 3}

 Let us find conditions for the characteristic algebra $L_x$  to be generated by
vector fields  $X$, $K$, $C_1$, $Z_1$ and $Z_2$. (We assume that $f_{t_xt_x}=0$.)

As in the previous cases to check that  $X$, $K$, $C_1$, $Z_1$ and $Z_2$ form a basis it is enough to check that $[C_1,Z_1]$ and $[K,Z_2]$ have unique expansion.
 Also we note that if  $[C_1,Z_1]$ and $[K,Z_2]$ can be expended with respect to  $X$, $K$, $C_1$, $Z_1$ and $Z_2$ then
\begin{equation}\label{case3.1}
[C_1,Z_1]=\bar \alpha Z_1,
\end{equation}
\begin{equation}\label{case3.2}
[K,Z_2]=\bar \lambda Z_1+\bar \mu Z_2
\end{equation}
for some functions $\bar\alpha,\,\bar\lambda$ and $\bar\mu$. This follows from the form of $[C_1,Z_1]$ and $[K,Z_2]$.
In general  one should write $[C_1,Z_1]=\bar \alpha Z_1 +\bar\beta Z_2$ but we show that $\bar\beta$ is zero in the next lemma.

\begin{lemma}
Let $f_{t_xt_x}=0$ then if the vector field $[C_1,Z_1]$ admits linear representation with respect to  vector fields  $X$, $K$, $C_1$, $Z_1$ and $Z_2$ then
equality Eq(\ref{case3.1}) holds.
\end{lemma}
\noindent
{\bf Proof.}
From the form of $[C_1,Z_1]$ it follows that $[C_1,Z_1]=\bar \alpha Z_1 +\bar\beta Z_2$. Let us show that $\bar \beta $ is zero.
We have $f_{t_xt_x}=0$ and  $f_{t_xt_x}=0$ if and only if
\begin{equation}\label{C2=0}
C_2=0.
\end{equation}
 Using definition of $Z_1$, $Z_2$ and Jacobi identity we have
\begin{equation}\label{[X,Z1]=0}
[X,Z_1]=[X,[K,C_1]]=-[K,[C_1,X]]-[C_1,[X,K]]=[K,C_2]-[C_1,C_1]=0
\end{equation}
 and
\begin{equation}
[X,Z_2]=[X,[K,Z_1]]=-[K,[Z_1,X]]-[Z_1,[X,K]]=[C_1,Z_1]
\end{equation}
Since $f_{t_xt_x}=0$ then $f_{t_x}$ does not depend on $t_x$ and coefficients of vector field
$$
C_1=\frac{\partial}{\partial t} + f_{t_x}\frac{\partial}{\partial t_1}+ g_{t_x}\frac{\partial}{\partial t_{-1}}+\dots
$$
do not depend on $t_x$.
The equality $[X,Z_1]=0$ implies that the coefficients of $Z_1$ also do not depend on $t_x$. Thus if $[C_1,Z_1]=\bar \alpha Z_1 +\bar\beta Z_2$
then functions $\bar\alpha$ and $\bar\beta$ do not depend on $t_x$, that is $X(\bar\alpha)=0$ and $X(\bar\beta)=0$.
Consider $[X,[C_1,Z_1]]$, from one hand, by Eq(\ref{C2=0}) and Eq.(\ref{[X,Z1]=0})
$$
[X,[C_1,Z_1]]=-[C_1,[Z_1,X]]-[Z_1,[X,C_1]]=-[C_1,[Z_1,X]]-[Z_1,C_2]=0,
$$
from the other hand,
$$
[X,[C_1,Z_1]]=[X,\bar\alpha Z_1+\bar\beta Z_2]=(X(\bar\alpha)+\bar\alpha\bar\beta)Z_1+(X(\bar\beta)+\bar\beta^2)Z_2=\bar\alpha\bar\beta Z_1+\bar\beta^2 Z_2
$$
Therefore, $\bar\alpha\bar\beta Z_1+\bar\beta^2 Z_2=0$ or $\bar\beta=0$. $\Box$

The next lemma shows that equalities Eq.(\ref{case3.1}) and Eq.(\ref{case3.2}) imply that   vector fields  $X$, $K$, $C_1$, $Z_1$ and $Z_2$  form a basis
of  $L_x$.

\begin{lemma}\label{case3_basis_lemma}
The vector fields $X$, $K$, $C_1$, $Z_1$ and $Z_2$ form a basis of the characteristic $x$-ring $L_x$ if and only if vectors fields $[C_1,Z_1]$ and  $[K,Z_2]$
admit a unique linear representations with respect to the basis vector fields.
\end{lemma}

\noindent
The above Lemma is proved in the same way as Lemma \ref{case1_basis_lemma}.

Let us write the systems corresponding to equalities  Eq.(\ref{case3.1}) and
Eq.(\ref{case3.2}).

\begin{lemma}\label{case3.1_lemma}
The equality Eq.(\ref{case3.1}) holds if and only if the $\bar\alpha$ and $(D\bar
\alpha)$ satisfy the following system
\begin{equation}
\label{case3.1.1_system}
\frac{1}{f_{t_x}}\bar\alpha-(D\bar\alpha)=\frac{f_{tt_x}+f_{t_x}f_{t_xt_1}}{f_{t_x}^2},
\end{equation}
\begin{equation}\label{case3.1.2_system}
(v-w)(D\bar\alpha)=\frac{f_{tt_x}+2f_{t_x}f_{t_xt_1}}{f_{t_x}^2}(w-v)+\frac{1}{f_{t_x}}C_1(v-w).
\end{equation}
\end{lemma}
{\bf Proof.}
Applying the automorphism $D(\cdot)D^{-1}$ to Eq.(\ref{case3.1}) we get
$$
D[C_1,Z_1]D^{-1}=(D\bar \alpha)D Z_1D^{-1},
$$
Direct calculations show that if $f_{t_xt_x}=0$ then
$$
D[C_1,Z_1]D^{-1}=\frac{1}{f_{t_x}^2}[C_1,Z_1]-\frac{f_{tt_x}+f_{t_x}f_{t_x
t_1}}{f_{t_x}^3}Z_1+\frac{1}{f_{t_x}^3}\left(\frac{f_{tt_x}+2f_{t_x}f_{t_xt_1}}{f_{t_x}}(w-v)+C_1(v-w)\right)C_1+...X
$$
Substituting the expressions for $D[C_1,Z_1]D^{-1}$, $DZ_1D^{-1}$  and comparing
coefficients before $C_1$ and $Z_1$  we obtain Eq.(\ref{case3.1.1_system}) and
Eq.(\ref{case3.1.2_system}) . $\Box$

\begin{lemma}
The equality Eq.(\ref{case3.2}) holds if and only if the coefficients $\bar
\mu$ and $\bar \lambda$ satisfy the following system
\begin{equation}\label{case3.2_system}
\begin{array}{lllllllll}
\bar E_{11}\bar \mu &+ & \bar E_{12}(D\, \bar\mu)& & & & &=& \bar F_1\\
& & \bar E_{22}(D\,\bar\mu)&+&\bar E_{23}\bar\lambda & +& \bar E_{24}(D\,
\bar\lambda)&=&\bar F_2\\
& &\bar E_{32}(D\,\bar\mu)&+& & &\bar E_{34}(D\, \bar\lambda)&=&\bar F_3\\
\end{array}
\end{equation}
where
$$
\bar E_{11}=1, \quad   \bar E_{12}=-1,   \quad \bar F_{1}=\frac{3w-v}{f_{t_x}},
$$
$$
\bar E_{22}=\frac{2w-v}{f_{t_x}},  \quad   \bar E_{23}=1,\quad  \bar  E_{24}= -1,
$$
\begin{eqnarray}\nonumber
 \bar F_{2}=p\bar
\alpha-f_{t_x}K\left(\frac{v-2w}{f_{t_x}^2}\right)-\frac{K(v-w)}{f_{t_x}}-\frac{2w(w-v)}{f_{t_x}^2}-
 \frac{f_t(v-w)}{f_{t_x}^2}-\frac{p}{f_{t_x}}(f_{tt_x}+f_{t_x}f_{t_xt_1}),
 \end{eqnarray}
$$
\bar
E_{32}=\frac{K(v-w)}{f_{t_x}^2}-\frac{2w(v-w)}{f_{t_x}^3}+\frac{f_t(v-w)}{f_{t_x}^3},
 \quad   \bar E_{34}=\frac{v-w}{f_{t_x}^2},
$$
\begin{eqnarray}\nonumber
\bar F_{3}=K\left(
\frac{K(v-w)}{f_{t_x}^2}+\frac{2w(w-v)}{f_{t_x}^3}+\frac{2f_t(v-w)}{f_{t_x}^3}\right)+pX\left(\frac{K(w-v)}{f_{t_x}^2}+\frac{2w(v-w)}{f_{t_x}^3}
\right)\\
+\frac{2pf_{tt_x}(w-v)-f_{t_x}^2Z_1(p)+f_{t_x}(w-v)C_1(p)+f_t(v-w)X(p)}{f_{t_x}^3}.
\end{eqnarray}
\end{lemma}
{\bf Proof.}
Applying the automorphism $D(\cdot)D^{-1}$ to Eq.(\ref{case3.2}) we get
$$
D[K,Z_2]D^{-1}=(D\bar\lambda)D Z_1D^{-1}+(D\bar \mu)D Z_2 D^{-1}.
$$
Direct calculations show that if $f_{t_xt_x}=0$ then
$$
DZ_2D^{-1}=\frac{1}{f_{t_x}}Z_2+\frac{v-2w}{f_{t_x}^2}Z_1+\frac{1}{f_{t_x}^3}\left(f_{t_x}K(v-w)-2w(v-w)+f_t(v-w)\right)C_1+...X
$$
and
$$
D[K,Z_2]D^{-1}=\frac{1}{f_{t_x}}[K,Z_2]+\frac{v-3w}{f_{t_x}^2}Z_2+TZ_1-\frac{p}{f_{t_x}}[X,Z_2]+RC_1+...X,
$$
where
$$
T=K\left(\frac{v-2w}{f_{t_x}^2}\right)+\frac{1}{f_{t_x}^3}(f_{t_x}K(v-w)-2w(v-w)+f_t(v-w)+pf_{t_x}(f_{tt_x}+f_{t_x}f_{t_xt_1})),
$$
$$
R=(K-pX)\left\{\frac{1}{f_{t_x}^3}(f_{t_x}K(v-w)-2w(v-w)+2f_t(v-w))\right\}
$$
$$-\frac{1}{f_{t_x}}Z_1(p)+\frac{w-v}{f_{t_x}^2}C_1(p)+\frac{f_t}{f_{t_x}^3}(v-w)X(p).
$$
Note that $[X,Z_2]=[C_1,Z_1]$.
Substituting the expressions for $D[K,Z_2]D^{-1}$, $DZ_1D^{-1}$, $DZ_2D^{-1}$ and
comparing coefficients of $C_1$, $Z_1$ and $Z_2$ we obtain Eq.(\ref{case3.2_system}).
$\Box$

\begin{theorem}
The characteristic $x$-ring of Eq.\ref{basic_eqn} is generated by vector
fields $X$, $K$, $C_1$, $Z_1$ and $Z_2$ if and only if the following conditions are
satisfied
  \begin{equation}\label{basic_system_cond1_case3}
  D\left(-f_{t_xt_x}+\frac{C_1(v-w)}{v-w}\right)=-\frac{f_{t_xt}+2f_{t_x}f_{t_xt_1}}{f_{t_x}^2}+\frac{C_1(v-w)}{f_{t_x}(v-w)},
  \end{equation},
 \begin{equation}\label{basic_system_cond2_case3}
\left(-\bar E_{11}\bar E_{22}\bar E_{34}(D^{-1}\bar E_{34})+\bar E_{11}\bar
E_{24}\bar E_{32}(D^{-1}\bar E_{34})- \bar E_{12}\bar E_{23}\bar E_{34}(D^{-1}\bar
E_{32})\right)\ne 0
 \end{equation}
  and
 \begin{equation}\label{basic_system_cond3_case3}
 (D\,\bar H)=\frac{\bar F_1}{\bar E_{12}}-\frac{\bar E_{11}}{\bar E_{12}}\bar H,
 \end{equation}
 where
 \begin{multline}
 \bar H=\left(\left(\bar F_1\bar E_{24}\bar E_{32}-\bar F_1\bar E_{22}\bar
E_{34}-\bar F_2\bar E_{12}\bar E_{34} -\bar F_3\bar E_{12}\bar
E_{24}\right)(D^{-1}\bar E_{34}) +(D^{-1}\bar F_{3})\bar E_{12}
 \bar E_{23}\bar E_{34}\right) \\
\left(-\bar E_{11}\bar E_{22}\bar E_{34}(D^{-1}\bar E_{34})+\bar E_{11}\bar
E_{24}\bar E_{32}(D^{-1}\bar E_{34})- \bar E_{12}\bar E_{23}\bar E_{34}(D^{-1}\bar
E_{32})\right)^{-1}
 \end{multline}
\end{theorem}
{\bf Proof.} In Lemma \ref{case3.1_lemma} we can easily find $\bar\alpha$ and
$(D\bar\alpha)$ independently. The condition that $(D\bar\alpha)$ is the shift of
$\bar\alpha$ leads to
Eq.(\ref{basic_system_cond1_case3}).
By Lemma \ref{gen_system_lem} the conditions Eq.(\ref{basic_system_cond2_case3}) and
Eq.(\ref{basic_system_cond3_case3}) imply that the system Eq.(\ref{case3.2_system}) have
unique solution.
Hence equalities  Eq.(\ref{case3.1}) and Eq.(\ref{case3.2}) hold and the characteristic
ring $L_x$ is generated by vector fields $X$, $K$, $C_1$, $Z_1$ and $Z_2$. $\Box$

\end{document}